\documentclass[12pt]{article}

\usepackage{amsmath}
\usepackage{amsfonts}
\usepackage{amsthm}
\usepackage{graphicx}
\usepackage{hyperref}
\usepackage{enumerate}
\usepackage{subcaption}
\usepackage[parfill]{parskip}

\numberwithin{equation}{section} 

\newtheorem{thm}{Theorem}[section]

\newtheorem{prop}[thm]{Proposition}

\begin{document}

\title{Solving satisfiability using inclusion-exclusion}

\author{
Anthony Zaleski\thanks{Department of Mathematics, Rutgers University (New Brunswick), 110 Frelinghuysen Road, Piscataway, NJ 08854-8019, USA.}
}

\maketitle
\begin{abstract}
Using Maple, we implement a SAT solver based on the principle of inclusion-exclusion and the Bonferroni inequalities.  Using randomly generated input, we investigate the performance of our solver as a function of the number of variables and number of clauses.  We also test it against Maple's built-in \verb+tautology+ procedure.  Finally, we implement the Lov\'asz local lemma with Maple and discuss its applicability to SAT.
\end{abstract}


\section{Introduction to SAT}
First, some terminology.  A \emph{Boolean variable} is a variable which can take on values in $\{true, false\}$, or, equivalently, $\{0,1\}$ (e.g. $x$).  A \emph{literal} is a Boolean variable or its negation (e.g. $\lnot x$).  \emph{Disjunction} means ``or" ($\lor$) and conjunction means ``and'' ($\land$).  A \emph{disjunctive clause} is a disjunction of literals (e.g. $x\lor \lnot y \lor z$); similarly, we can define the \emph{conjunctive clause}. A \emph{conjunctive normal form} (CNF) is a conjunction of disjunctive clauses (e.g. $\lnot z \land (y \lor  z)\land(x\lor \lnot y)$); similarly, we can define the \emph{disjunctive normal form} (DNF).  

We say that a CNF $S$ in the variables $x_1,\dots,x_n$ is \emph{satisfiable} iff there exists an assignment of truth values to $x_1,\dots,x_n$ that makes $S$ true.  For example, the CNF in the previous paragraph is  satisfiable: the first clause forces $z=false$; then the second forces $y=true$; and the third forces $x=true$, giving us a valid assignment.  On the other hand, the CNF $(x \lor y) \land \lnot x \land \lnot y$ is, of course, not satisfiable.  
 
Given a CNF in $n$ variables, one obvious way to determine its satisfiability is to check all $2^n$ assignments to the variables.  There is an ongoing effort to develop more efficient algorithms to determine satisfiability.  We call these algorithms ``SAT solvers.''  Currently, even the most efficient SAT solvers are exponential time; one can always construct worst-case scenarios that take long for the algorithm to analyze.  In fact, SAT has been shown to be NP-complete, so a polynomial time SAT solver would indeed be breaking news.

Here, we shall certainly not present a polynomial-time algorithm, or even one that is practically more competent than current solvers.  Rather, we wish to outline a simple, novel approach to solving SAT, analyze its strengths and weaknesses, and discuss how it might be used as the basis for a more powerful solver.
\section{SAT and inclusion-exclusion}
Suppose $S=C_1\land\cdots\land C_N$ is a CNF with $N$ clauses and $n$ variables $x_1,\dots,x_n$.  Then, $S$ is satisfiable iff $\lnot S=\lnot C_1 \lor \cdots \lor \lnot C_N$ is not a tautology.  So SAT can be rephrased as ``given an arbitrary DNF, determine if it is a tautology.''  We shall use this formulation in our approach.

Thus, let $S=C_1\lor\cdots\lor C_N$ be a DNF with $N$ clauses and $n$ variables $x_1,\dots,x_n$.  We wish to determine if all $2^n$ possible assignments to the variables result in $S$ being true.  We can interpret this probabilistically: If we pick a uniform random assignment, is $\Pr[S=true]=1$? Equivalently, letting $A_k$ be the event that $C_k$ is satisfied, is $\Pr[\cup_k A_k]=1$?

Recall that we can compute the probability of such a union using the following:
\begin{prop}[Principle of Inclusion-Exclusion]
Let $A_1,\dots,A_N$ be events in a finite probability space.  For $I \subset [N]$, define $$A_I=\bigcap_{j\in I}A_j.$$  Then,
$$
\Pr[\cup_k A_k]=\sum_{i=1}^N (-1)^{i+1} \sum_{I \subset [N], |I|=i} \Pr[A_I].
$$
\end{prop}

So our problem amounts to finding $\Pr[A_I]$ for arbitrary $I \subset [N]$, which is easy: Let $V$ be the set of literals appearing in the clauses $\{C_j : j \in I\}$; then, $\Pr[A_I]=0$ if $V$ contains a variable and its negation, and $\Pr[A_I]=2^{-|V|}$ otherwise.

This idea is easily implemented to produce a simple SAT solver which always terminates with a correct answer.  Such a solver, along with some test results, is briefly outlined in [GC].

However, notice that the sums in Proposition 2.1 grow with the number of clauses.  Luckily, we have the \emph{Bonferroni inequalities}, which tell us that we can compute the outer sum partially and still get a bound on the probability we are after:
\begin{prop}[Bonferroni Inequalities]
With the notation of Proposition 2.1, let $1\leq k \leq N$.  Then,
$$
\Pr[\cup_k A_k] \bowtie \sum_{i=1}^k (-1)^{i+1} \sum_{I \subset [N], |I|=i} \Pr[A_I],
$$
where $\bowtie$ means $\leq$ if $k$ is odd and $\geq$ if $k$ is even.
\end{prop}

Using the Bonferroni inequalities and looping over $k$, we can get a sequence of upper and lower bounds on $\Pr[\cup_k A_k]$.  If at some point we find that $\Pr[\cup_k A_k]<1$ or $\Pr[\cup_k A_k]\geq 1$, then we can exit the loop and determine that $S$ is not a tautology or a tautology, respectively.  In the worst case, we have to go up to $k=N$, but (hopefully) we arrive at a decision after significantly less steps.
\subsection{Details of the algorithm}
The method outlined above is implemented in the Maple package \verb+sat.txt+; see Section 5 for instructions to obtain the package.  

We encode a DNF as a set of sets of integers: For example, \verb+{{1,-2},{3}}+ corresponds to $(x_1 \land \lnot x_2) \lor x_3$. The \verb+Merge+ procedure is the equivalent of conjunction: \verb+Merge({-1,2},{2,3})+ returns \verb+{-1,2,3}+, while  \verb+Merge({1,2},{-2,3})+ returns \verb+false+ since these two clauses are ``incompatible,'' i.e., not simultaneously satisfiable.

The main procedure is \verb+Taut+.  It inputs a DNF \verb+S+ and threshold \verb+K+. We initialize \verb+P=0+ and \verb+N=nops(S)+, the number of clauses.  For \verb+k+ from 1 to \verb+K+, we compute the \verb+k+th term in the inclusion-exclusion sum and add it to \verb+P+.  For the sake of efficiency, a table is used to keep track of all \emph{compatible} conjunctions of \verb+k+ clauses in \verb+S+, so that at the \verb+k+th stage, the table has at \emph{most} 
\verb+N+ choose \verb+k+ entries.  If we obtain a conclusive bound at some point in the loop, we return \verb+[ans,k]+, where the first entry is \verb+true+ or \verb+false+, depending on whether we found \verb+S+ to be a tautology.  If we complete the whole loop without coming to a conclusion, we return \verb+[P,k]+.
\section{Testing the solver}
To test our solver, we use the procedure \verb+RandNF(n,N,M)+, which generates a random DNF with \verb+N+ clauses in \verb+n+ variables, each containing \verb+M+ uniform random literals.  By default, \verb+M=3+, which we shall assume from now on.

The procedure \verb+MetaTaut(n,N,K,M)+ runs \verb+Taut+ on \verb+M+ random DNFs with \verb+n+ variables and \verb+N+ clauses and threshold \verb+K+, and it records the run time and output of each trial.

The procedure  \verb+MetaTaut(n,N,K,M)+  does the same, but instead of our solver, it uses Maple's built-in \verb+tautology+ procedure.
\subsection{Runtimes}
As one would expect, our solver seems to perform most competently when there are lots of variables but not too many clauses.  

For example, Figure 1 shows a histogram of runtimes resulting from using \verb+Taut+ on 1000 random DNFs generated by \verb+RandNF(100,10)+. In all of these cases, our solver arrived at the correct answer by the third step of the loop, and the longest runtime was .006s. As Figure 1 shows, the Maple solver performed slower in this case.  

Further, we tested \verb+Taut+ on 10 random DNFs generated by \verb+RandNF(1000,20)+, and it decided each of them was not a tautology by the seventh inclusion-exclusion step.  The runtimes ranged from 2-58 minutes, with an average of 19.  In this case, using \verb+MapleTaut+ resulted in an overflow error.

On the other hand, Figure 2 shows the results when 100 random DNFs generated by \verb+RandTaut(100,20)+ are used.  Already, the number of clauses is enough to make our solver slower than Maple.  In fact, in this case, only fifteen of the 100 random DNFs are solvable by \verb+Taut+ with threshold $k=6$.

Also, we should point out that, in the situations where our method does seem promising, it seems that it almost always returns false.  So, as it is, it probably has little practical use.  Further, we are only testing it against a na\:ive built-in Maple tautology function, rather than a sophisticated SAT solver. 
\begin{figure}
\centering
\begin{subfigure}[l]{0.45\textwidth}
\includegraphics[width=\textwidth]{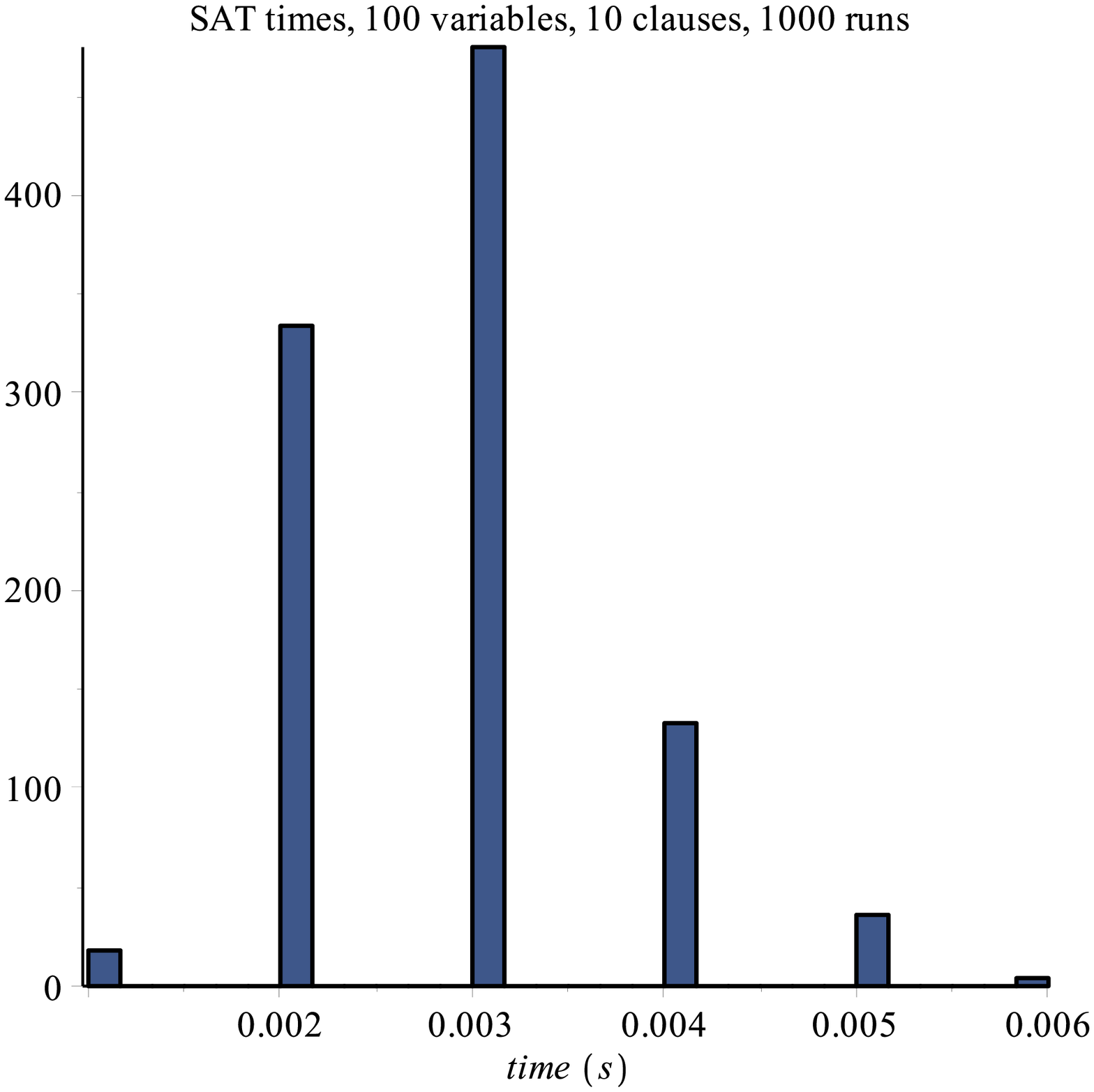}
\end{subfigure}
\begin{subfigure}[r]{0.45\textwidth}
\includegraphics[width=\textwidth]{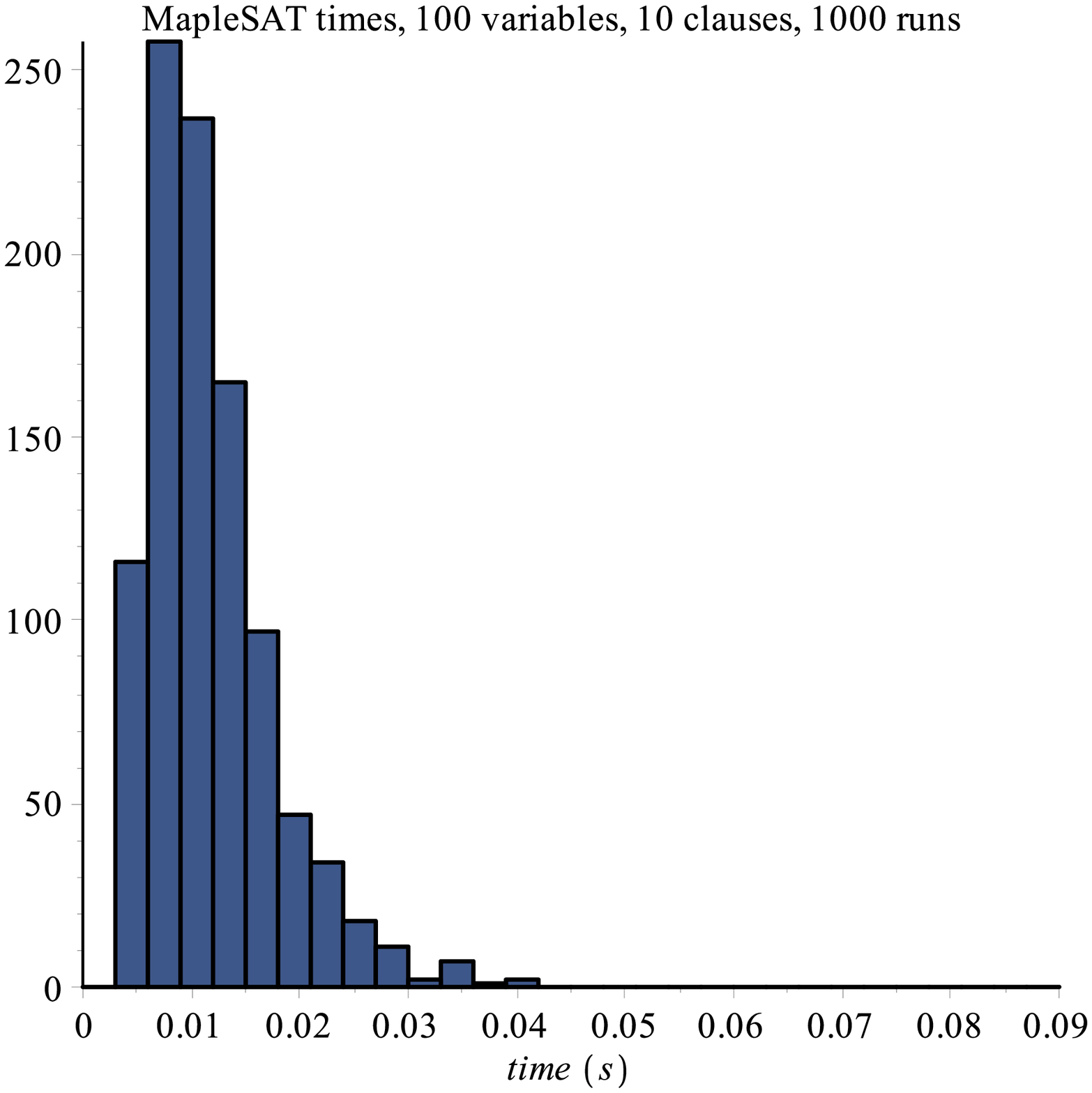}
\end{subfigure}
\caption{As shown in these runtime frequency plots, when the variable to clause ratio is high enough, our solver (left) out-performs Maple (right).}
\end{figure}
\begin{figure}
\centering
\begin{subfigure}[l]{0.45\textwidth}
\includegraphics[width=\textwidth]{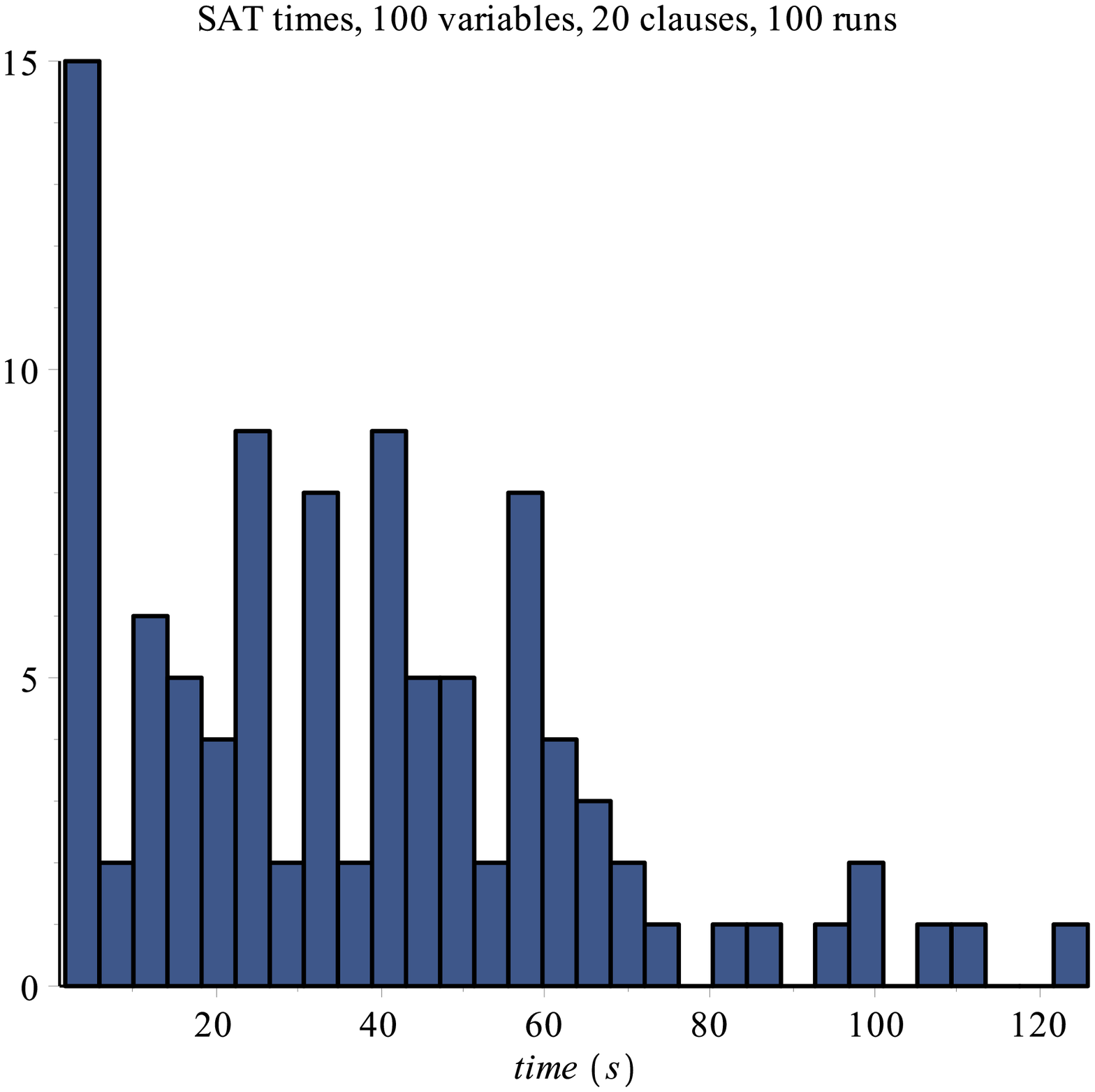}
\end{subfigure}
\begin{subfigure}[r]{0.45\textwidth}
\includegraphics[width=\textwidth]{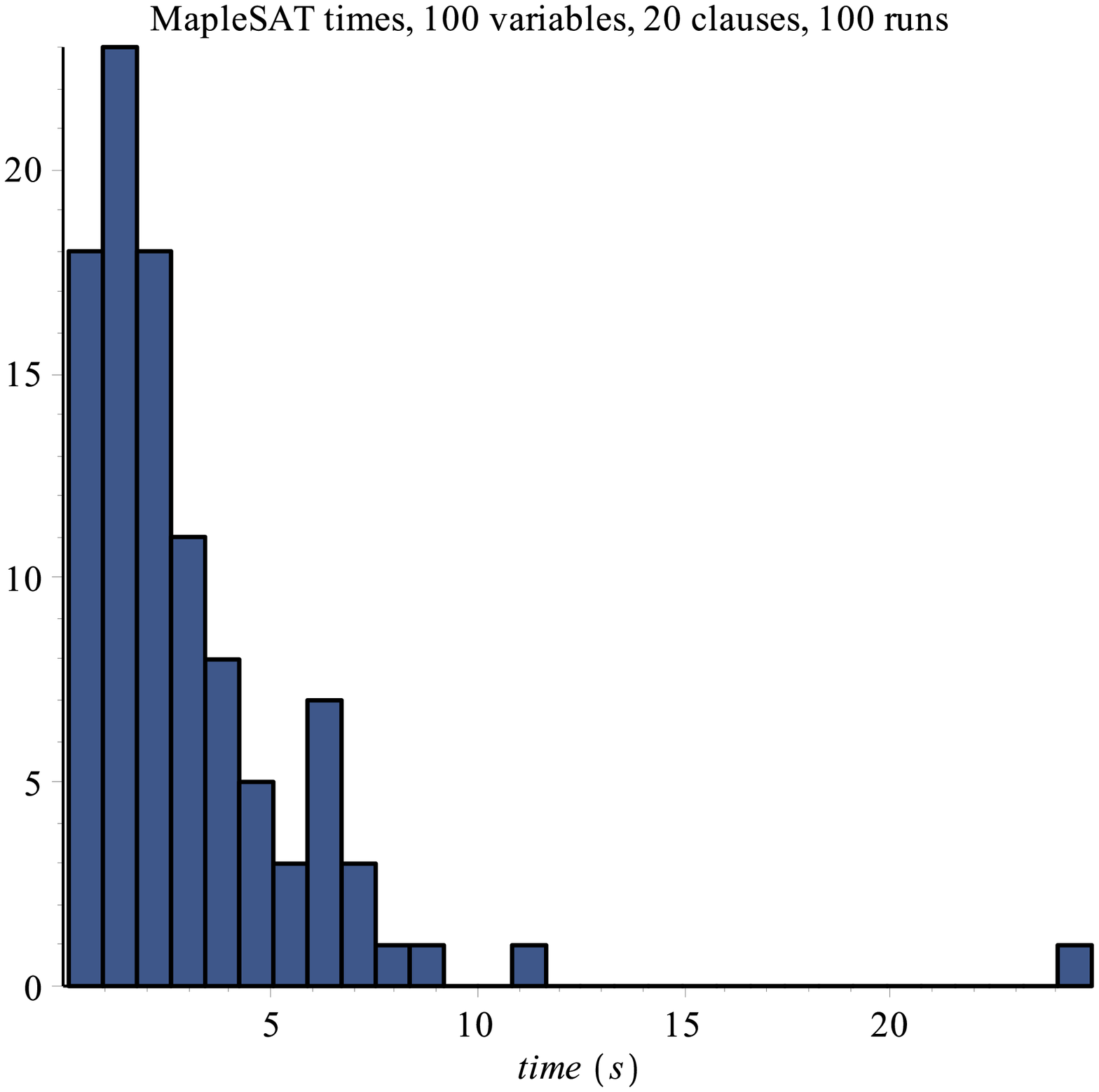}
\end{subfigure}
\caption{With a lower variable-to-clause ratio, our solver (left) loses to Maple (right).}
\end{figure}
\subsection{Thresholds}
Recall that, in \verb+Taut(S,k)+, the argument \verb+k+ is the threshold, that is, the number of inclusion-exclusion summands computed before the procedure quits.  Now, we investigate how the required threshold is related to the number of variables \verb+n+ and number of clauses \verb+N+.

The procedure \verb+HowManyFinished(n,N,k,M)+ runs \verb+Taut+ with threshold \verb+k+ on \verb+M+ random DNFs generated by \verb+RandNF(n,N)+, and it outputs the proportion of conclusive runs.  In other words, it estimates \emph{success probability} that a DNF generated by \verb+RandNF(n,N)+ is solvable by our algorithm with threshold \verb+k+.

Empirical evidence shows a phase shift behavior in the success probability If we fix $n$ and $k$ and vary $N$.  Namely, there appears to be a critical number of clauses $N_c(n,k)$ at which the graph of the success probability has an inflection point.  Of course, we have $N_c>k$, with $N_c$ increasing in $k$.

Some plots exhibiting this phase shift are shown in Figure 3.  Note that this behavior is reminiscent of the satisfiability phase shift studied in [XW], where the behavior of the probability of a random CNF being satisfiable as a function of the ratio of the number of variables and clauses is studied.

\begin{figure}
\centering
\includegraphics[width=.7\textwidth]{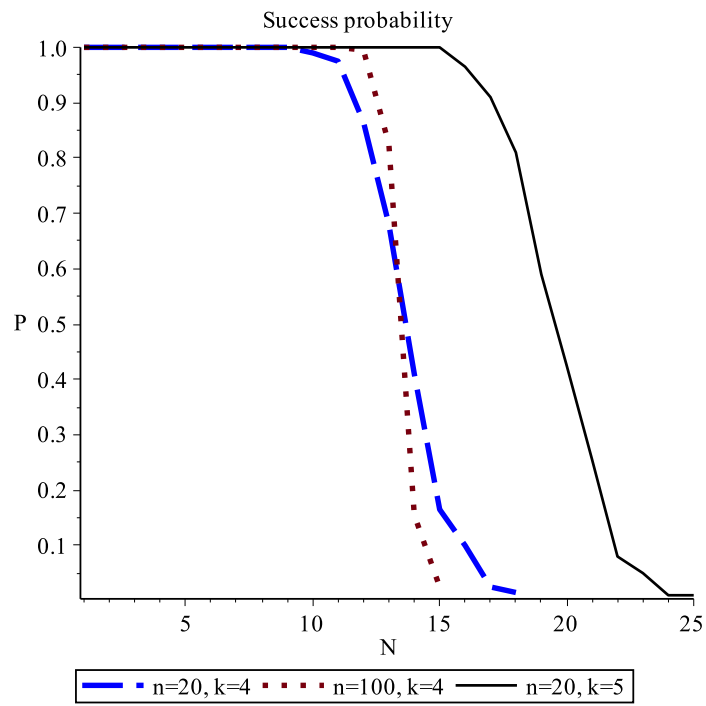}
\caption{Here, $n$ and $N$ correlate with the number of variables and clauses, respectively; $k$ is the threshold used in our solver; and $P$ is the proportion of times our solver was successful, based on 200 runs with random DNFs.}
\end{figure}

\section{SAT and the Lov\'asz local lemma}
\subsection{Computerizing the local lemma}
Given some ``bad events'' $\mathcal{A}=\{A_1,\dots,A_N\}$, the Lov\'asz local lemma can be used to verify that there is a positive probability that \emph{none} of them occurs.  Suppose $G$ is a \emph{dependency graph} on the vertex set $\mathcal{A}$: That is, events in $\mathcal{A}$ are mutually independent of their non-neighbors in $G$.  Let $\Gamma(A)$ denote the neighborhood of $A$ in $G$.  Then the following holds:

\begin{prop}[Asymmetric Lov\'asz local lemma]
Suppose there exists a weight function $x: \mathcal{A} \to [0,1)$ such that
$$
\forall A \in \mathcal{A}, \,\,\,\, \Pr(A)\leq x(A) \prod_{B \in \Gamma(A)} (1-x(B)).
$$
Then $\Pr(\bigcap_i A_i^c)>0$.
\end{prop}

In applications, the weight function $x(A)$ is usually found ad hoc.  If we assume each vertex of the dependency graph has degree $\leq d$ and set $x\equiv 1/(d+1)$, we obtain the following:

\begin{prop}[Symmetric Lov\'asz local lemma]
Suppose each event $A_i$ satisfies $\Pr(A_i)\leq p$ and is independent of all but at most $d$ of the other events.  If
$$
ep(d+1)\leq 1,
$$
then $\Pr(\bigcap_i A_i^c)>0$.
\end{prop}

The procedure \verb+LLLs(P,G)+ in the Maple package checks if the events $A_i$ satisfy the symmetric local lemma, where the lists $P$ and $G$ satisfy$P[i]=\Pr(A_i)$ and $G[i]=\{j : A_j \in \Gamma(A_i)\}$.

Computerizing the asymmetric local lemma is harder, since, as far as we know, there is no systematic and efficient way to look for a valid weight function $x$.  Somewhat arbitrarily, the procedure \verb+LLL(P,G)+ uses the weight function $x(A)=1/(|\Gamma(A)|+1)$.  The motivation for this choice is that, when the dependency graph is uniform, it reduces to the symmetric local lemma.
\subsection{Applying the local lemma to SAT}

The article [G] addresses a theoretical application of the local lemma to SAT, focusing on using it to derive combinatorial conditions for the satisfiability of CNFs.  Here, present a computer application of the local lemma to SAT.

Let us return to the setup in Section 2.  We have a DNF $S=C_1\lor\cdots\lor C_N$ with variables $x_1,\dots,x_n$, which are assigned true/false values uniformly at random.  We let $A_k$ be the event that $C_k$ is true.  Then $S$ is \emph{not} a tautology iff there is a positive probability that \emph{none} of the events $A_k$ occurs.  So we can apply the local lemma.  

We form a dependency graph $G$ by  joining $A_i$ and $A_j$ iff the clauses $C_i$ and $C_j$ have common variables.  Also, $\Pr(A_i)=2^{-n_i}$, where $n_i$ is the number of literals in $C_i$; for example, for 3-SAT, $\Pr(A_i)=1/8$.

The procedure \verb+DNFtoPG(S)+ converts the DNF $S$ to a pair $P,G$, which can be passed to one of the \verb+LLL+ procedures.  If the procedure returns \emph{true}, then we can conclude that $S$ was \emph{not} a tautology; otherwise, this method is inconclusive.

Unfortunately, \verb+LLLs+ rarely succeeds at detecting a non-tautology,  and \verb+LLL+ is only slightly better.  For example, out of 100 non-tautologies generated by \verb+RandNF(100,10)+,  only 26 were detected  by \verb+LLLs+ and 37 by \verb+LLL+. Out of 100 non-tautologies generated by \verb+RandNF(100,15)+,  only 2 were detected  by \verb+LLLs+ and 3 by \verb+LLL+. We expect that this is due to the behavior of the dependency graph.  It would be interesting to develop a ``clever'' asymmetric local lemma algorithm that tailors the weight function to work for the given dependency graph.
\section{Using the Maple package}
The Maple package \verb+sat.txt+ accompanying this paper may be found at the following URL:
 \\ \url{http://www.math.rutgers.edu/~az202/Z}.  
 
 To use the Maple package, place \verb+sat.txt+ in the working directory and execute \verb+read(`sat.txt`);+.  
 
 To see the main procedures, execute \verb+Help();+.  For help on a specific procedure, use \verb+Help(<procedure name>);+.  
 
 Here are some things to try:
 \begin{itemize}
 \item \verb+Taut({{-3,1,4},{-2,-1,4}},2);+ determines if $(\lnot x_3 \land x_1 \land x_4)\lor (\lnot x_2 \land \lnot x_1 \land x_2)$ is a tautology using inclusion-exclusion with threshold 2.
 \item  \verb+MapleTaut({{-3,1,4},{-2,-1,4}});+ determines if $(\lnot x_3 \land x_1 \land x_4)\lor (\lnot x_2 \land \lnot x_1 \land x_2)$ is a tautology using Maple's \verb+tautology+ procedure.
 \item \verb+RandNF(10,4);+ generates a random DNF with approximately 10 variables and 4 clauses.
 \item \verb+MetaTaut(5,25,8,10);+ runs \verb+Taut(RandNF(5,25),8)+ 10 times.
 \item \verb+LLL(DNFtoPG({{-3,1,4},{-2,-1,4}}));+ determines if the LLL applies to the given DNF (if true is returned, then it is not a tautology).
 \item \verb+MetaLLL(100,10,100);+ applies the local lemma to 100 DNFs generated by \verb+RandNF(100,10)+ and outputs the proportion of non-tautologies detected by \verb+LLL+ and the actual proportion of non-tautologies.
\end{itemize}
\section*{Acknowledgement} 
The author thanks Dr. Doron Zeilberger for introducing this project to him
and guiding his research in the right direction.
\section{References}
\begin{itemize}
\item[{[G]}] Heidi Gebauer, Robin A. Moser, Dominik Scheder, and Emo Welzl, The Lov\'asz Local Lemma and Satisfiability,  in: Efficient Algorithms, Part I, pp. 30-54, Susanne Albers, Helmut Alt, and Stefan N\:aher (eds.).  Heidelberg: Springer-Verlag, 2009.
\item[{[GC]}]
G\'abor Kusper and Csaba Bir\'o, Solving SAT by an Iterative Version of the Inclusion-Exclusion Principle, 2015 17th International Symposium on Symbolic and Numeric Algorithms for Scientific Computing (SYNASC), Timisoara, 2015, pp. 189-190.
\item[{[XW]}]
Ke Xu and Wei Li, The SAT phase transition, 
\\
\url{http://www.math.ucsd.edu/~sbuss/CourseWeb/Math268_2007WS/satphase.pdf}
\end{itemize}

\end{document}